# Moment Transform-Based Compressive Sensing in Image Processing

Theofanis Kalampokas and George A. Papakostas

HUman-MAchines INteraction Laboratory (HUMAIN-Lab), Dept. of Computer Science, International Hellenic University Kavala, Greece
{theokala, gpapak}@cs.ihu.gr

**Abstract.** Over the last decades, images have become an important source of information in many domains, thus their high quality has become necessary to acquire better information. One of the important issues that arise is image denoising, which means recovering a signal from inaccurately and/or partially measured samples. This interpretation is highly correlated to the compressive sensing theory, which is a revolutionary technology and implies that if a signal is sparse then the original signal can be obtained from a few measured values, which are much less, than the ones suggested by other used theories like Shannon's sampling theories. A strong factor in Compressive Sensing (CS) theory to achieve the sparsest solution and the noise removal from the corrupted image is the selection of the basis dictionary. In this paper, Discrete Cosine Transform (DCT) and moment transform (Tchebichef, Krawtchouk) are compared in order to achieve image denoising of Gaussian additive white noise based on compressive sensing and sparse approximation theory. The experimental results revealed that the basis dictionaries constructed by the moment transform perform competitively to the traditional DCT. The latter transform shows a higher PSNR of 30.82 dB and the same 0.91 SSIM value as the Tchebichef transform. Moreover, from the sparsity point of view, Krawtchouk moments provide approximately 20-30% more sparse results than DCT.

## 1 Introduction

Compressive Sensing is an important contribution for the reason that it overthrows the traditional sampling theory of Shannon and Nyquist where in order to reconstruct a signal without error the sampling rate should be at least twice the signal's maximum frequency component. Compressive sensing as introduced by Donoho et al. [1], Candes et al. [2] gave the potential to reconstruct a full signal even if it is sparse. With these results over the past decades has been done a significant effort in the development of techniques in many domains around CS and sparse representation theory because it affects many domains from the perspective of storage to the recovery of corrupted information despite the structure of the signal. With the introduction of this theory many image processing and computer vision problems are affected not only in the compression but to denoising or inpainting of images. Besides CS-based denoising or inpainting, there is a variety of research work that has been done in image watermarking [3], image hiding [4].

In general CS and sparse representation theory makes two conjectures: that any natural image can be sparse in a basis or a dictionary, constructed by Fourier, wavelet,

DCT, or any other transforms, where only a few transform coefficients are significant and the rest are zero or negligible, and that the measurement basis is incoherent with the basis, which the image is sparse [5].

In image denoising, the techniques can be divided into two groups: *spatial-domain* and *transform-domain* methods. In the former category are included methods based on the Perona-Malik equation, which brings good noise reduction and edges preservation [6], or the bilateral filter technique, which acquires the original image with noise reduction [7]. The transform-domain denoising methods include some modern algorithms arising from CS and sparse representation theory where the sparsity of the signal is exploited through some transform. For example, Strack et al.[8] proposed a new geometric multiscale transform Curvelet and in comparison with wavelet, it achieved better results in image denoising.

There is a huge variety of implementations that have been proposed around the above theory, from the different transformations in the basis dictionary to the creations of new algorithms that are applied to several data structures for different domains. In this paper, the moment transform is proposed in CS and sparse representation theory for the description of images in a denoising task, in comparison with classic transformation Discrete Cosine Transform (DCT). More specifically, two different moment families are used the Tchebichef and Krawtchouk moments. The former moment family is known for its noise invariance and later one for its capabilities to describe a signal locally.

The main contribution of this paper is the proposal for the first time, the moment transform in the denoising problem based on CS and sparse representation theory, towards advantaging from the noise tolerance and local description of the examined moment families. From the experimental results, it is proven that moment transform brings more sparse results and thus more compressible than DCT according to the increase of noise level, although brings similar reconstruction quality.

The rest of this paper is organized as follows: Section 2 presents the theory of transform-based compressive sensing. Section 3 describes the moment transform, with emphasis on the Tchebishef and Krawtchouk moments used in this work, along with the proposed methodology. The proposed moment-based compressive sensing methodology is described in Section 4 and experimental results are discussed in Section 5. Finally, Section 6 concludes this work.

## 2  Transform-based Compressive Sensing

The sampling technique in CS theory gives the potential to sample a signal at rates proportional to the amount of information in the signal by exploiting the sparsity properties of signals. Considering that a signal e.g. an image or can have a sparse representation when expressed w.r.t. some basis or a dictionary $\boldsymbol{\Psi} \in \mathbb{R}^{n \times m}$. This dictionary shows a complete structure where $n = m$, or an overcomplete structure where $n < m$. Then a signal $\boldsymbol{x}$ can be represented as a linear combination of basis $\boldsymbol{\Psi}$ atoms, with $\boldsymbol{a} \in \mathbb{R}^{n \times m}$ being the projection coefficients of signal $\boldsymbol{x}$ in $\boldsymbol{\Psi}$ domain, expressed in as:

$$\boldsymbol{x} = \boldsymbol{\Psi}\boldsymbol{a}, \qquad (1)$$

then the compressed signal $y$ is derived as follows:

$$y = \Phi x = \Phi \Psi a = Da. \tag{2}$$

If the processed signal is contaminated with noise or it is corrupted with irrelevant content (2) can be expressed as:

$$y = Da + z. \tag{3}$$

In (3) $z$ represents noise measurements of any type and $D$ is the sensing matrix. The meaning of compression is that signal $y$ will be smaller than the original signal x. Thus the task is to recover $x$ based on $y$, $D$, and $\Psi$. The sensing matrix $D$ guarantees the signal sparsity or the separation between noise and signal content and must satisfy the Restricted Isometry Property (RIP) for any k-sparse signal $x$ as proposed by [9] and expressed as:

$$(1 - \delta)\|a\|^2 \leq \|Da\|^2 \leq (1 + \delta)\|a\|^2, \tag{4}$$

where $\delta_k \in (0, 1)$ is the Restricted Isometry Constant (RIC).

The above theory highlights the importance of the transformation selection that will describe the image in a more sparse form. The intrinsic properties of the signals may not be suitably interpreted by all transformations, which makes sparsity not feasible. The transformation selection as the basis dictionary of the signal for different tasks is a topic that still concerns the research community. Ansari et al. [10] implemented a comparison between various transformations e.g. Wavelet, Curvelet, and Contourlet for denoising remote-sensed images contaminated with Gaussian noise. They concluded that Curvelet denoising preserves better the sharpness of the boundaries. Starck et al. [11] proposed Undecimated Wavelet Transform (UWT) where it is proven that overcomes the disadvantage of the discrete wavelet transform regarding its shift invariance property. Wang et al. [12] proposed Shearlet Transform (ST) in CS theory, which is a directional multiresolution transformation, providing higher PSNR in different sampling ratios than the Wavelet Transform (WT). The shifting of the input signal causes small changes to the transform coefficients, which results in a bad representation of edges and borders. Dragotti et al. [13] introduced Directionlets, which is a transformation that provides an efficient interpretation for the nonlinear approximation of images and compared to the WT provides better PSNR with similar complexity.

## 3  Moment Transform

Image moments are the coefficients of the Moment Transform (MT) that has achieved a significant contribution around signal processing in a variety of domains [14]. Their advantage arises from the robustness in describing an image with fewer coefficients than the actual size of the image, which is a characteristic that matches the CS and sparse representation theory. Among several moment types, the orthogonal moments include orthogonal polynomials as kernel functions, owing desirable properties in both continuous and discrete coordinate spaces [15]. Due to the orthogonality property, image moments provide a more compact representation of an image and robustness to noise.

The first orthogonal moments were expressed in continuous space and it was Zernike, Pseudo-Zernike, Fourier-Mellin, Legendre and are used in many applications for feature description as Kadir et al. [16]. The disadvantage of continuous orthogonal moments is the approximation errors that arise for the reason of the coordinate normalization and space granulation procedures. To overcome this disadvantage discrete orthogonal moments have proposed where defined inside the discrete coordinate system of the image, with the Tchebichef [17], Krawtchouk [18], and dual Hahn [19] moments being the most representative discrete moment families. As a transformation image moments have been proposed in many applications from pattern recognition [20] and adversarial computer vision [21] to the interpretation of EEG signals for seizure classification [22] where the proposed method is proven the robustness of the moment transform in the presence of noise.

Tchebichef moments are robust to high noise levels and object description in comparison with other image orthogonal moments. The Tchebichef Moments (TMs) for an image with $NxN$ pixels size are defined as:

$$T_{nm} = \frac{1}{\bar{\rho}(n,N)\bar{\rho}(m,N)} \sum_{x=0}^{N-1} \sum_{y=0}^{N-1} \tilde{t}_n(x)\tilde{t}_m(y)f(x,y). \tag{5}$$

In (5) the first term $\bar{\rho}(n,K)$ corresponds to the normalized norm of the Tchebichef polynomials and the term $\tilde{t}_n(x)$ is the normalized Tchebichef polynomials defined as:

$$\tilde{t}_n(x) = t_n(i)/\beta(n,N), \tag{6}$$

with

$$\begin{aligned}t_n(i) &= (1-N)_n 3F_2(-n,-x,1+n;1,1-N;1) \\ &= n! \sum_{k=0}^{n} (-1)^{n-k} \binom{N-1-k}{n-k}\binom{n+k}{n}\binom{x}{k},\end{aligned} \tag{7}$$

and $\beta(n,N)$ is usually equal to $N^n$.

Krawtchouk moments [18] are another family of discrete orthogonal moments, characterized by their high local representation capabilities. The locality characteristic of Krawtchouk moments is controlled by the parameters $p_1, p_2$ and express the spread of the coefficient calculation in an image. The Krawtchouk moments (KMs) of order $n$ and repetition $m$ for a $NxN$ pixels are computed with the following formula:

$$K_{nm} = \sum_{x=0}^{N-1} \sum_{y=0}^{N-1} \bar{K}_n(x;p_1,N-1)\bar{K}_m(y;p_2,N-1)f(x,y), \tag{8}$$

where $\bar{K}_m$ are the weighted Krawtchouk polynomials defined in [18].

## 4   Compressive Sensing based on Moment Transform

As discussed above, image moments are robust to noise presence and describe the content of an image in a compact way. Based on the theory, there are two alternative strategies for denoising: one is by *compression*, which is achieved through the transformation where in this occasion is the *moment transform* and the second is the optimal sparse representation of the signal through the transformation. The computed basis dictionary for Tchebichef and Krawtchouk moments, which are examined in this study can be expressed as:

$$\Phi_{nm} = [Poly_n]^T Poly_m, \qquad (9)$$

where $Poly_n$ and $Poly_m$ are the $n^{th}$ and $m^{th}$ order orthogonal polynomials of any moment family, respectively. According to (9) the basis dictionary of a specific moment family is extracted in order to be used in sparse coding or reconstruction algorithm to recover an image.

Next, the reconstruction process takes place through a sparse recovery algorithm from the family of greedy algorithms, which are faster than convex algorithms. These algorithms rely on an iterative approximation of the image coefficients, by obtaining an improved estimation of sparse representation of the image at each iteration that attempts to account for the mismatch to the measured data. The algorithm that is used in this work is the Orthogonal Matching Pursuit (OMP) [23], which is an improvement of the matching pursuit algorithm. It computes the inner product of the residue and the measurement matrix and then selects the index of the maximum correlation column, extracts this column in each iteration and adds it to the selected set of atoms. Then an orthogonal projection is performed over the subspace of previously selected atoms, which provides a new approximation vector used to update the residual. Since in the measurement matrix the columns are orthogonal there will be no column selected twice.

As presented in (2) there is an attempt to select the fewer columns of $D$ that participate in $y$. With OMP as a reconstruction algorithm, there are two steps that need to be fulfilled according to:

$$\lambda_t = arg \max_{j=1,2\dots,N} |\langle r_{t-1}, \Psi_j \rangle|, \qquad (10)$$

where $\lambda_t$ is the index that solves the above optimization problem, which consists of the terms $r$ and $\Psi$ which are the residual and the transformation basis, respectively. Next, a least squares problem is solved to obtain a new image estimation:

$$x_t = arg \min_x \|D_t x - y\|_2. \qquad (11)$$

Then a new approximation of data and the new residual are calculated:

$$a_t = D_t x_t, \;(13)\; r_t = y - a_t. \qquad (12)$$

The above are calculated for a number ($t$) of iterations and terminates when a stopping criterion is met.

## 5 Experiments

In this paper, the denoising of five benchmark images *Lena, Barbara, Baboon, Pirate, and Peppers*, contaminated with Gaussian additive noise, using sparse coding with OMP algorithm is applied, and Tchebichef, Krawtchouk, and DCT basis dictionaries are examined. The experiments are implemented in Python with the usage of Scikit-learn library and executed in a Ryzen 3700 CPU with 16GB RAM. Figures 1-3, depict the basis dictionaries for each transformation.

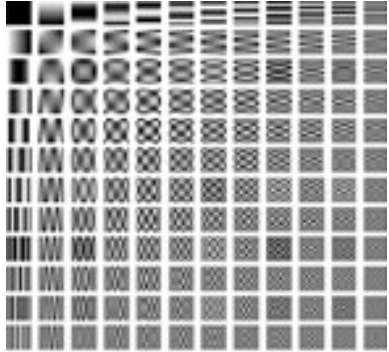  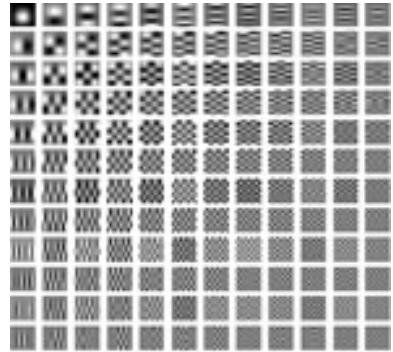

**Fig. 1.** Tchebichef basis dictionary     **Fig. 2.** Krawtchouk basis dictionary

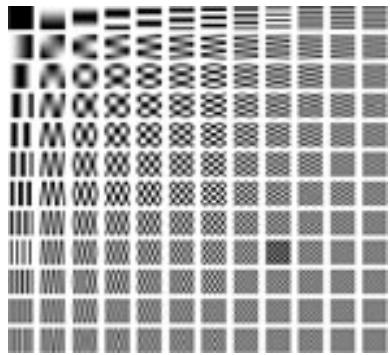

**Fig. 3.** DCT basis dictionary

Next, the input images are described in overlapped patches of 12x12 pixels size in all image dimensions, in order to achieve a redundant structure between image information and noise presence. The selection of this patch dimension is calculated by the square root of image dimensions, which are 144x144. Such patches are presented as 144-dimensional vectors of gray-scale values. The DC value and the mean of the gray-scale values were subtracted from each vector as a preprocessing step. The result is a linear dependency between the components of the observed data and therefore the dimension of the data was reduced by one. The denoising results for each benchmark image and different noise ratios are presented in the following Table 1-5 and Fig. 4-8. It is worth noting that the PSNR (in dB) and SSIM indices are used to evaluate the quality of the denoised images. First the mathematical expression of PSNR is:

$$\text{PSNR} = 10\log_{10}\frac{N^2}{\sum_{i=0}^{N-1}\sum_{j=0}^{N-1}\left(x_{i,j}^{(p)}-x_{i,j}^{(0)}\right)^2}. \quad (13)$$

In (13) N is the maximum value of the pixel and the denominator is the reconstruction Mean Squared Error. The definition of SSIM is presented as:

$$\text{SSIM}(x,y) = \frac{(2\mu_x\mu_y + c_1)(2\sigma_{xy} + c_2)}{(\mu_x^2 + \mu_y^2 + c_1)(\sigma_x^2 + \sigma_y^2 + c_2)}. \quad (14)$$

In (14) x and y are the two images and $\mu_x$ with $\mu_y$ are the average of each image, $\sigma_x^2$ with $\sigma_y^2$ is the variance of each image, $\sigma_{xy}$ is the covariance of the two images and finally $c_1$ and $c_2$ are two variables in order to stabilize the operation in the occasion that the denominator is weak.

**Table 1.** TMs, KMs, DCT denoising performance in Peppers.

| Noise Ratio | PSNR-TMs | PSNR-KMs | PSNR-DCT | SSIM-TMs | SSIM-KMs | SSIM-DCT |
|---|---|---|---|---|---|---|
| 0.1 | 27.27 | 27.02 | 26.92 | 0.91 | 0.88 | 0.91 |
| 0.2 | 24.35 | 25.36 | 23.96 | 0.78 | 0.79 | 0.78 |
| 0.3 | 22.32 | 23.54 | 21.81 | 0.67 | 0.69 | 0.66 |
| 0.4 | 20.16 | 21.48 | 19.64 | 0.57 | 0.58 | 0.56 |
| 0.5 | 18.21 | 19.52 | 17.98 | 0.48 | 0.49 | 0.47 |

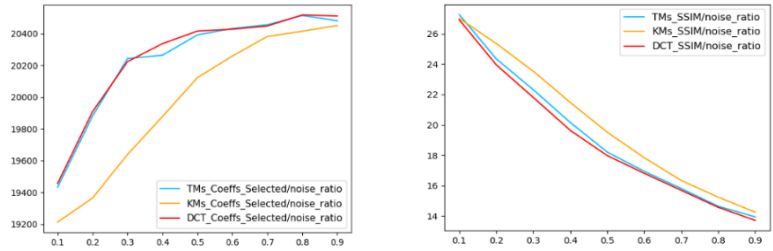

**Fig. 4.** Selected coefficients on the left and SSIM on the right for TMs, KMs and DCT in Peppers.

**Table 2.** TMs, KMs, DCT denoising performance in Barbara.

| Noise Ratio | PSNR-TMs | PSNR-KMs | PSNR-DCT | SSIM-TMs | SSIM-KMs | SSIM-DCT |
|---|---|---|---|---|---|---|
| 0.1 | 27.69 | 26.51 | 27.96 | 0.88 | 0.85 | 0.88 |
| 0.2 | 26.19 | 25.47 | 26.36 | 0.81 | 0.79 | 0.81 |
| 0.3 | 24.26 | 23.82 | 24.29 | 0.72 | 0.71 | 0.72 |
| 0.4 | 21.68 | 21.84 | 21.71 | 0.61 | 0.62 | 0.62 |
| 0.5 | 20.38 | 20.13 | 19.86 | 0.54 | 0.53 | 0.53 |

8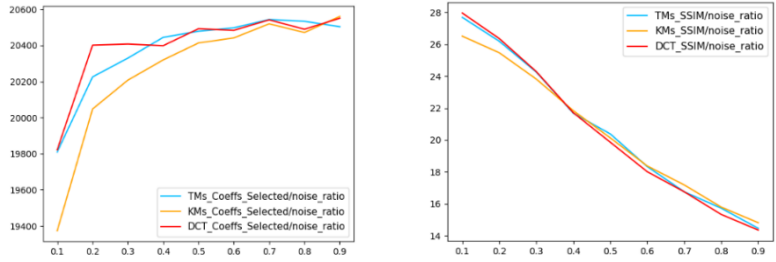

**Fig. 5.** Selected coefficients on the left and SSIM on the right for TMs, KMs and DCT in Barbara.

**Table 3.** TMs, KMs, DCT denoising performance in Lena.

| Noise Ratio | PSNR- TMs | PSNR- KMs | PSNR-DCT | SSIM- TMs | SSIM- KMs | SSIM-DCT |
|---|---|---|---|---|---|---|
| 0.1 | 30.47 | 29.84 | 30.82 | 0.91 | 0.89 | 0.91 |
| 0.2 | 28.04 | 27.38 | 28.08 | 0.81 | 0.79 | 0.79 |
| 0.3 | 25.22 | 24.87 | 25.07 | 0.68 | 0.67 | 0.66 |
| 0.4 | 22.88 | 22.93 | 22.89 | 0.57 | 0.58 | 0.57 |
| 0.5 | 20.75 | 20.99 | 20.74 | 0.48 | 0.49 | 0.48 |

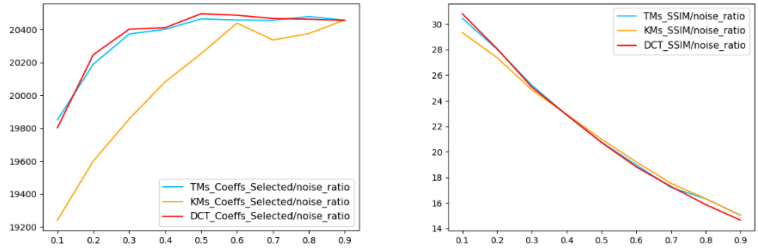

**Fig. 6.** Selected coefficients on the left and SSIM on the right for TMs, KMs and DCT in Lena.

**Table 4.** TMs, KMs, DCT denoising performance in Baboon.

| Noise Ratio | PSNR- TMs | PSNR- KMs | PSNR-DCT | SSIM- TMs | SSIM- KMs | SSIM-DCT |
|---|---|---|---|---|---|---|
| 0.1 | 24.89 | 24.23 | 24.84 | 0.79 | 0.76 | 0.79 |
| 0.2 | 24.03 | 23.47 | 23.96 | 0.73 | 0.71 | 0.73 |
| 0.3 | 22.74 | 22.42 | 22.63 | 0.65 | 0.65 | 0.65 |
| 0.4 | 21.33 | 21.55 | 21.21 | 0.57 | 0.56 | 0.57 |
| 0.5 | 19.98 | 19.89 | 19.88 | 0.51 | 0.49 | 0.51 |

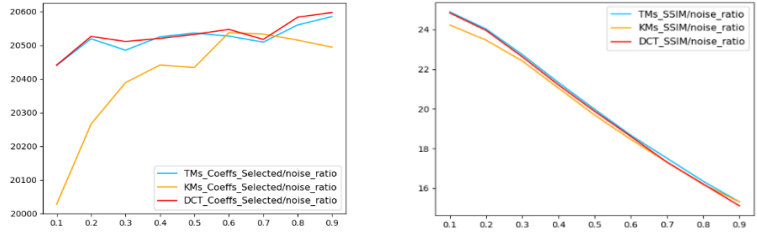

**Fig. 7.** Selected coefficients on the left and SSIM on the right for TMs, KMs and DCT in Baboon.

**Table 5.** TMs, KMs, DCT denoising performance in Pirate.

| Noise ratio | PSNR-TMs | PSNR-KMs | PSNR-DCT | SSIM-TMS | SSIM-KMs | SSIM-DCT |
|---|---|---|---|---|---|---|
| 0.1 | 27.01 | 26.16 | 27.04 | 0.85 | 0.82 | 0.85 |
| 0.2 | 25.66 | 25.21 | 25.68 | 0.77 | 0.76 | 0.77 |
| 0.3 | 23.87 | 23.72 | 23.18 | 0.69 | 0.68 | 0.68 |
| 0.4 | 21.98 | 22.03 | 21.81 | 0.58 | 0.59 | 0.58 |
| 0.5 | 20.07 | 20.25 | 19.83 | 0.48 | 0.51 | 0.49 |

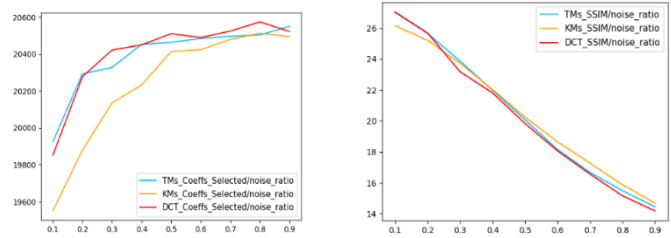

**Fig. 8.** Selected coefficients on the left and SSIM on the right for TMs, KMs and DCT in Pirate.

From the above results, it is concluded that moment transform in comparison with DCT is bringing more sparse results and almost equal PSNR and SSIM metrics. TMs and DCT perform almost equal for both PSNR and SSIM, in the case of Peppers, Baboon, and Pirate images, but TMs achieve more sparse results than DCT. Next in Barbara and Lena DCT achieves the highest PSNR in low noise ratios with small differences with TMs, where it takes the leading place for higher noise ratio than DCT, a conclusion that is confirmed in all cases. Same as previous in the last two cases TMs achieve more sparse results than DCT. Finally, an important notice is that TMs are more robust than DCT as the noise level increases.

Krawtchouk moments and DCT achieve the sparsest solutions in all cases. KMs are not achieved high scores in PSNR and SSIM like DCT in low noise ratio but we can come to the conclusion that same with TMs for higher noise ratios KMs are bringing almost equal metrics with DCT and even better in some cases. Finally, it is important to mention that according to all figures, image moments in comparison with DCT are bringing better metrics for higher noise ratios and the sparsest solutions in all levels of noise for all benchmark images. In the following Fig. 9-13, the denoised images for the case of 20% noise ratio for all benchmark images are presented.

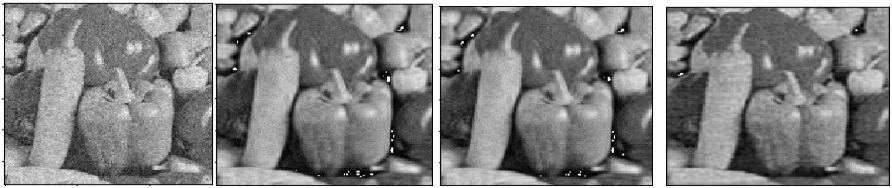

**Fig. 9.** Peppers: From left to right noisy image, DCT, TMs, and KMS denoised images.

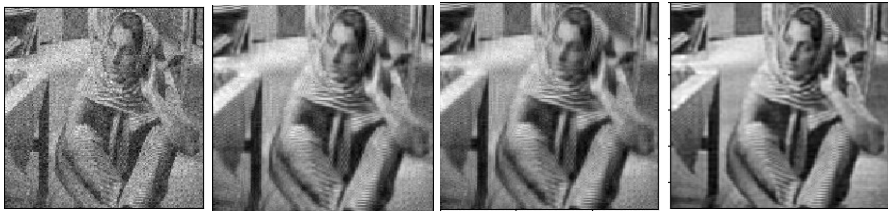

**Fig. 10.** Barbara: From left to right noisy image, DCT, TMs, and KMS denoised images.

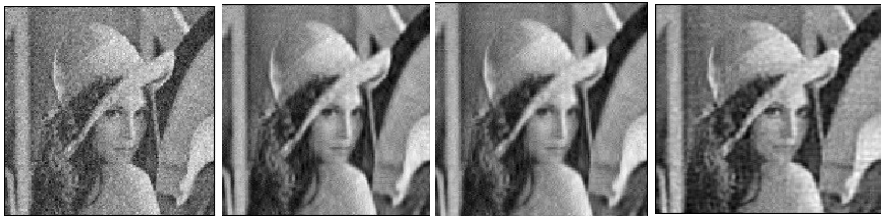

**Fig. 11.** Lena: From left to right noisy image, DCT, TMs, and KMS denoised images.

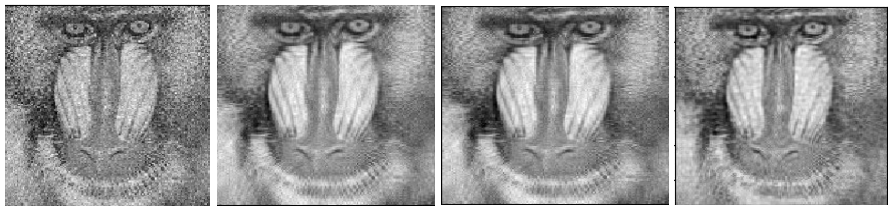

**Fig. 12.** Baboon: From left to right noisy image, DCT, TMs, and KMS denoised images.

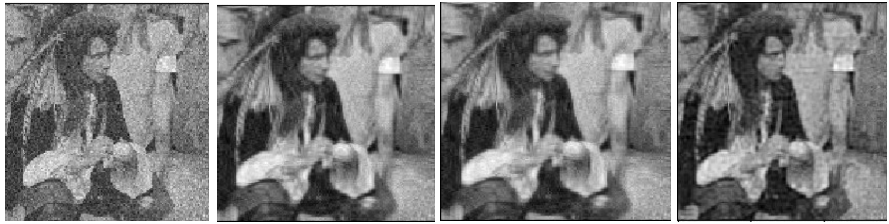

**Fig. 13.** Pirate: From left to right noisy image, DCT, TMs, and KMS denoised images.

## 6  Conclusions

The previous study proved that moment transform based on Tchebichef and Krawtchouk families are bringing more sparse solutions than DCT with almost equal metrics. As a result, image moments are suitable for Compressive Sensing and sparse representation theory in denoising applications from the point of strong and convenient feature invariances. Based on the above conclusion the interest arises on what other problems of this theory can moment transform be applied from dictionary learning to other domains. Another conclusion is that other types of image moments could be applied like fractional or quaternion moments recently proposed in the literature.

**Acknowledgements**
This work was supported by the MPhil program "Advanced Technologies in Informatics and Computers", hosted by the Department of Computer Science, International Hellenic University, Greece.

## References

1. Donoho, D. L. (2006). Compressed sensing. *IEEE Transactions on Information Theory*, *52*(4), 1289–1306. https://doi.org/10.1109/TIT.2006.871582
2. Candès, E. J. (2006). Compressive sampling. International Congress of Mathematicians, ICM 2006, 3, 1433–1452. https://doi.org/10.4171/022-3/69
3. Boulanger, S., Mitchell, G., Bouarab, K., Marsault, É., Cantin, A., Frost, E. H., & Déziel, E. (2009). A compressive-sensing based watermarking schme for sparse image tampering identification. Politecnico di Milano Dip. Elettronica e Informazione - Italy Universit ` a di Siena Dip . Ingegneria dell ' Informazione - Italy. System, 59(12), 1265–1268.
4. Hua, G., Xiang, Y., & Bi, G. (2016). When compressive sensing meets data hiding. *IEEE Signal Processing Letters*, *23*(4), 473–477. https://doi.org/10.1109/LSP.2016.2536110
5. Candès, E., & Romberg, J. (2007). Sparsity and incoherence in compressive sampling. *Inverse Problems*, *23*(3), 969–985. https://doi.org/10.1088/0266-5611/23/3/008
6. Guo, Z., Sun, J., Zhang, D., & Wu, B. (2012). Adaptive perona-malik model based on the variable exponent for image denoising. *IEEE Transactions on Image Processing*, *21*(3), 958–967. https://doi.org/10.1109/TIP.2011.2169272
7. Elad, M. (2002). On the origin of the bilateral filter and ways to improve it. *IEEE Transactions on Image Processing*, *11*(10), 1141–1151. https://doi.org/10.1109/TIP.2002.801126

8. Starck, J. L., Candès, E. J., & Donoho, D. L. (2002). The curvelet transform for image denoising. *IEEE Transactions on Image Processing*, *11*(6), 670–684. https://doi.org/10.1109/TIP.2002.1014998
9. Candes, E. J., & Tao, T. (2005). Decoding by linear programming. *IEEE Transactions on Information Theory*, *51*(12), 4203–4215. https://doi.org/10.1109/TIT.2005.858979
10. Ansari, R. A., & Budhhiraju, K. M. (2016). A Comparative Evaluation of Denoising of Remotely Sensed Images Using Wavelet, Curvelet and Contourlet Transforms. *Journal of the Indian Society of Remote Sensing*, *44*(6), 843–853. https://doi.org/10.1007/s12524-016-0552-y
11. Starck, J. L., Fadili, J., & Murtagh, F. (2007). The undecimated wavelet decomposition and its reconstruction. *IEEE Transactions on Image Processing*, *16*(2), 297–309. https://doi.org/10.1109/TIP.2006.887733
12. Wang, F., Wang, S., Hu, X., & Deng, C. (2012). Compressive sensing of image reconstruction based on shearlet transform. *Advances in Intelligent and Soft Computing*, *125 AISC*, 445–451. https://doi.org/10.1007/978-3-642-27329-2_61
13. Dragotti, P. L. (2006). Directionlets : Anisotropic Multidirectional. *Image (Rochester, N.Y.)*, *15*(7), 1916–1933.
14. Papakostas, G. A. (2015). Improving the recognition performance of moment features by selection. In: Stańczyk U., Jain L. (eds) Feature Selection for Data and Pattern Recognition. Studies in Computational Intelligence, 584, 305–327. https://doi.org/10.1007/978-3-662-45620-0_13
15. Papakostas, G.A. (2014). Over 50 Years of Image Moments and Moment Invariants. In: Papakostas G.A. (ed) Moments and Moment Invariants - Theory and Applications. Gate to Computer Science and Research (GCSR), 1, 3–32. https://doi.org/10.15579/gcsr.vol1.ch1
16. Kadir, A., Nugroho, L. E., Susanto, A., & Insap Santosa, P. (2012). Experiments of zernike moments for leaf identification. *Journal of Theoretical and Applied Information Technology*, *41*(1), 82–93.
17. Mukundan, R., Ong, S. H., & Lee, P. A. (2001). Image analysis by Tchebichef moments. *IEEE Transactions on Image Processing*, *10*(9), 1357–1364. https://doi.org/10.1109/83.941859
18. Yap, P. T., Paramesran, R., & Ong, S. H. (2003). Image analysis by Krawtchouk moments. *IEEE Transactions on Image Processing*, *12*(11), 1367–1377. https://doi.org/10.1109/TIP.2003.818019
19. Zhu, H., Shu, H., Zhou, J., Luo, L., & Coatrieux, J. L. (2007). Image analysis by discrete orthogonal dual Hahn moments. *Pattern Recognition Letters*, *28*(13), 1688–1704. https://doi.org/10.1016/j.patrec.2007.04.013
20. Papakostas, G. A., Koulouriotis, D. E., Karakasis, E. G., & Tourassis, V. D. (2013). Moment-based local binary patterns: A novel descriptor for invariant pattern recognition applications. *Neurocomputing*, *99*(January), 358–371. https://doi.org/10.1016/j.neucom.2012.06.031
21. Maliamanis, T. & Papakostas, G. A. (2021). DOME-T: adversarial computer vision attack on deep learning models based on Tchebichef image moments. Proc. SPIE 11605, Thirteenth International Conference on Machine Vision, 116050D (4 January 2021). https://doi.org/10.1117/12.2587268
22. Tziridis, K., Kalampokas, T., & Papakostas, G. A. (2021). EEG Signal Analysis for Seizure Detection Using Recurrence Plots and Tchebichef Moments. IEEE 11th Annual Computing and Communication Workshop and Conference (CCWC). (in print)

Tropp, J. A., & Gilbert, A. C. (2007). Signal Recovery From Random Measurements Via Orthogonal Matching Pursuit. 53 (12), 4655–4666.